\documentclass[12pt]{iopart}
\usepackage{graphicx}
\usepackage{dcolumn}
\usepackage{bm}
\begin{document}
\begin{center}
\underline{{\it Pramana -- Journal of Physics} {\bf (2010) (in press)}}
\end{center}
\title{Locating Phase Transitions in Computationally Hard Problems}
\author{B Ashok$^1$\footnote{Corresponding Author; Email: basp@uohyd.ernet.in, bashok1@gmail.com} and T K Patra$^2$\footnote{Current address: Dept. of Chemical Engineering, Indian Institute of Technology (IIT), Kanpur, India.}}
\address{$^1$ Advanced Centre of Research in High Energy Materials (ACRHEM),
University of Hyderabad, Central University PO, GachiBowli, Hyderabad - 500 046, India.}
\address{$^2$ School of Physics, University of Hyderabad, Central University PO,
GachiBowli, Hyderabad - 500 046, India.}

\begin{abstract}
We discuss how phase-transitions may be detected in computationally hard
problems in the context of Anytime Algorithms. Treating the computational time,
value and utility functions involved in the search results 
in analogy with quantities in statistical physics, we indicate how the onset of
a computationally hard regime can be detected and the transit to higher quality
solutions be quantified by an appropriate response function. The existence of a
dynamical critical exponent is shown, enabling one to predict the onset of
critical slowing down, rather than finding it after the event, in the specific
case of a Travelling Salesman Problem. This can be used as a means of improving
efficiency and speed in searches, and avoiding needless computations.\\

\noindent {\bf Keywords\/}: New applications of statistical mechanics, analysis
of algorithms, heuristics, phase transitions and critical phenomena
\end{abstract}
\pacs{05.70.Fh,~89.75.-k,~64.60.Bd,~64.60.A-}
\maketitle
\section*{\large{\bf 1. Introduction}}
\vspace*{-0.1cm}
\indent
Everyday computational situations are replete with cases where the solution quality does not seem
to justify the time spent in generating the answer, while a quick even if less accurate
solution would have been as satisfactory.
In the following sections, we describe how detection of phase transitions
in computational problems may be done, within a statistical mechanics
framework, thereby avoiding regimes wherein computational efforts put in are
not commensurate with the solution obtained. This would be of utmost importance
especially in situations where the problem being solved is an NP (nondeterministic
polynomial time) hard problem, or where there is a transition in the very nature
of the problem from an easily computable regime to one that is very hard to solve.
To be able to actually predict a phase-transition in a computational context
(not just find it subsequently) and to predict and avoid the ``critical slowing down
regime" has not, to the best of our knowledge, been demonstrated before.\\
\indent
We discuss this in detail in the context of Anytime Algorithms.
This framework has been completely explained in this paper itself,
so that this work is self-contained and complete. We have implemented this method here,
yielding very interesting and significant results.
There is a considerable body of literature dealing with the travelling salesman problem and with
phase transitions \& their use in constraint satisfaction problems
~\cite{kirkpatrick83}-\cite{hartmannweigt}.
While the use of phase transitions in search problems in general has been
extensively investigated, there does not seem to have been an attempt in
trying to use phase transitions in anytime algorithms as a means of
monitoring the progress of the algorithm or in any other form.
We show that a critical exponent can be found that relates the time taken to come to a
stable solution quality preceding a phase transition, to the number of nodes or cities
of a travelling salesman problem. We believe this is a new and important result that
has not been reported in the literature, as far as we know. The existence
of a critical exponent makes it possible to predict when the onset of a computationally
hard regime would take place. This is of great importance as it has long been unclear
how to efficiently run algorithms especially when one has to make a trade-off between
solution quality and the computational time needed to achieve an acceptable, near-optimal
solution, given constraints of time.\\
\indent
The particular problem used -- of the Travelling Salesman Problem,
is only illustrative of the theoretical approach we suggest for detecting
phase transitions. We show how this relatively simple
approach may be used in any other search and optimization
problem so that computational time may be vastly reduced by halting the running
of an algorithm when a near-optimal solution is reached prior to a phase transition
to a computationally hard regime.
It would be of interest to see if phase transitions could be used in conjunction
with anytime algorithms as a means of further improving
accuracy and speed in searches.
\section*{\large{\bf 2. A brief sketch of anytime algorithms}}
\vspace*{-0.1cm}
Anytime algorithms are algorithms whose execution can be stopped at any time,
giving a solution to the problem being solved. The quality of the solution
improves with time~\cite{shlomoAI,hansenshlomo}.
These algorithms become specially important when we are faced with constraints
on our resources--which might include, amongst other things, the computational
resources and the time available to us to solve the problem.
Thus, to satisfy the constraints put on it, an anytime algorithm trades-off
between the solution quality and the computational time, for example.
Two main types of anytime algorithms are contract and interruptible algorithms.
Contract algorithms run for a fixed length of time, and give a solution only
when that interval of time elapses. Interruptible algorithms form a larger
family -- they may be interrupted at any stage to obtain an answer though the
accuracy and  the meaningfulness of the solution would, of course, vary.
Monitoring an anytime algorithm is of great importance as we can keep track
of the progress of the algorithm and decide when and at what intervals we
could monitor or stop the algorithm in order to obtain an optimal answer
optimally.\\
\indent
There are some quantities which play a very basic part in such algorithms.
A performance profile gives us a measure of the expected quality of the
output with the execution time.
A performance profile is thus typically a probability distribution.
Those performance profiles which we have tacitly used are actually dynamic
performance profiles, $Pr(Q_j{\mid}{Q_i,\Delta t})$, which is the probability
of obtaining a solution of quality $Q_j$ by resuming the algorithm for
time interval $\Delta t$ when the current solution has quality $Q_i$.
A Utility function $U(Q_i,t_k)$ tells us the utility of a solution
of quality $Q_i$ at time $t_k$~\cite{hansenshlomo}.
In the work of Hansen and Zilberstein~\cite{hansenshlomo} a monitoring policy $m$ is
chosen for tracking the progress of an algorithm such that it maximises a value
function $V(f_r,t_k)$, a cost function $C_1$ being introduced to include the cost
of monitoring.
As one can only estimate the true solution quality at
any given time, a feature $f_r$ is made the basis for estimating the solution quality $Q_i$.
Use is made of partial observability functions like $Pr(Q_i{\mid}{f_r,t_k})$
and $Pr(f_r{\mid}{Q_i,t_k})$  in the value function above to estimate
improvements in quality. The algorithm is accordingly monitored or allowed to run or halted,
as dictated by the monitoring policy. Our approach is somewhat different, as explained in the
next section.
\section*{\large{\bf 3. Phase transitions and Anytime Algorithms}}
Although phase transitions occur only in the thermodynamic limit, in the
limit of infinitely large systems, as was first pointed out by Kramers~\cite{kramers},
finite state transitions nevertheless exhibit similarities
to true phase transitions and hence are very much relevant to computational
situations. Singularities and singular behaviour could be used to define universality
classes on the basis of common characteristics like, for example, swiftly changing
correlation lengths between parts of a system at and near the point of transition.\\

Phase transitions have appeared in a number of computational contexts and
paradigms (see, for example ~\cite{hogg87,hogg96,hartmannweigt}, and references therein).
The transition from a polynomial to exponential search costs, the transition
from an underconstrained to overconstrained problem, the appearance of
transitions in optimization problems, automatic planning and models of
associative memory amongst others, all indicate the widespread prevalence of
phase transitions in a computational context.
It has also been found that on an average several search heuristics have hard
problem instances concentrated at similar parametric values which points
correspond to transitions in solubility.\\
\indent
When one wishes to employ an anytime algorithm for getting a quick, approximate solution
to a problem, the question that arises is how long is long enough before stopping
the execution of the algorithm and deciding to accept a solution?
Can one decide beforehand, on a systematic basis, the length of this run-time?
For a problem, the time of onset of a transition to a computationally hard regime
wherein the solution quality does not significantly improve enough to justify running
the algorithm for a longer time, could act as one criterion for deciding the run-time.
Detecting such a transition requires one to run the algorithm unmindful of solution
quality so that any change in behaviour that quantifies the transition can be located.
This is what we have done in this paper.\\
\indent
Another requirement is to describe the particular computational problem being considered
using some more quantifiable terms.
Several authors have addressed combinatorial problems through ideas from statistical mechanics;
a good review of the literature can be found in~\cite{hartmannweigt} and references therein.
The work of Gent and Walsh~\cite{gentwalsh} addresses, in particular, the problem of a phase transition
for the Travelling Salesman Problem. They identify a transition between soluble and insoluble
instances of the decision problem (namely, whether or not a tour of some length $l$ or less exists for
the given TSP), at some critical value of a parameter.\\
\indent
In physical systems, a transition
from fluid to gas states, or from paramagnetic to ferromagnetic states for magnetic systems,
can be well described using the behaviour of the thermodynamical potentials and
response functions like susceptibility and order parameters like the
magnetization (see, for example, Reference~\cite{stanley}).\\
In the following sections, we present a formalism for quantifying phase transitions in the computational
context.\\
The test-problem that we study in this paper is that of obtaining a near-optimal solution for a
two-dimensional Travelling Salesman Problem (TSP) with $N$ nodes or cities distributed randomly over a
unit square. The premise is that any change in behaviour of the computational time of the algorithm
would be detected, enabling us to quantify any transition to a computationally hard regime.\\
 The challenge that we face here is how to draw an analogy between thermodynamic potentials and
response functions for a physical system to quantities in a more abstract system such as a
computational problem and algorithm.
\section*{\large{\bf 4. Defining the formalism in analogy with statistical physics}}
In this paper, we consider the two-dimensional Travelling Salesman Problem as the toy-problem to test
our formalism on.
In keeping with expectations that any phase transition in a system should be distinctly observable,
we first decide therefore to look for a drastic transition in various quantities like the Quality of a
solution and the Value function. The actual utility of a solution at any time-instant depends upon
its quality, as well as the rate of improvement shown by the quality and the computational time needed
to arrive at it. The utility can be expected to increase in direct proportion to both the quality of
the solution as well as the rate of improvement of quality, while reducing in direct proportion to the
increasing computational time required. We therefore make an ansatz that a Utility function denoted by
$U(Q_i,{\dot Q}_i,t_k)$ can be defined with these properties, and which in the simplest instance
has a linear dependence on $Q$ and ${\dot Q}$:
\begin{equation}
U(Q_i,{\dot Q}_i,t_k) = a_1 Q_i + a_2 {\dot Q}_i - a_3 t .
\end{equation}
We have taken $a_1$, $a_2$ and $a_3$ to be positive constants. The Quality function $Q_i$ can be defined
in several ways, depending upon the particular problem being considered. For the TSP, one obvious
choice for $Q_i$ would be the reduction in path-length over each iteration, since one wishes to
complete a tour as optimally as possible.
We hence choose to define quality by
\begin{equation}
Q_i \equiv (L_i - L_c)/L_i,
\end{equation}
$L_i$ being the initial path length and $L_c$ the current path length. The value function (mentioned earlier in Section 2)
used is taken as a sum of the expected values of the utility function, and is defined by
\begin{equation}
V(f_k,t_k) = {\sum}_j~Pr(Q_j{\mid}{f_k,t_k,{\Delta}t})U(Q_j,{\dot Q}_j,t_k + \Delta t) .
\end{equation}
In the toy model being discussed, we will take the feature $f_k$ to be the quality $Q_k$, so that $V(Q_k,t_k)$, defines in
quantitative terms the value of a solution of quality $Q_k$ at time $t_k$ in obtaining
other solutions of quality $Q_j$ in a time step following $t_k$.
The utility $U(Q_j,{\dot Q}_j,t_k + \Delta t)$ when weighted by the conditional probability
of obtaining that solution of quality $Q_j$ in time step $\Delta t$ after starting at time $t_k$
with quality $Q_k$, and summed over all $Q_j$, clearly gives a measure of the value of the
solution of quality $Q_k$. Hence the definition (3) of the value function.\\
\indent
The procedure adopted is as follows. 2-Opt heuristics is employed to solve each problem instance.
To calculate and generate a performance profile, the algorithm is run for about 500 times. To then
solve a particular problem, the performance profile generated earlier is used on completion of
each iteration of the algorithm to calculate quantities like the Value function. The initial tour
is generated by means of a nearest-neighbour algorithm. The steps of the initial tour construction
are:
\begin{enumerate}
\item Select a random city.
\item Find the nearest unvisited city and go there.
\item If any unvisited cities are left, repeat the previous step.
\item Return to the first city.
\end{enumerate}
Tour improvement is done through a 2-Opt heuristic; i.e., following initial tour construction,
in subsequent iterations, the algorithm selects a small segment of the sequence and reverses it.
We use a simple 2-bond move which reverses the sequence of cities for a chosen segment to obtain
a trial tour. In each iteration of this 2-opt heuristic, we accept a tour if it is shorter in length
than the previous one, rejecting it otherwise. Solution quality therefore keeps improving or remains
unchanged, as long as time permits. On completion of each iteration of our
algorithm, we use the performance profile generated earlier to calculate the value function and
the cost inflection function described below. Data from 500 samples are averaged for obtaining
each point in the data sets used to draw our inferences.\\
\vspace*{0.5cm}\\
\includegraphics[height=6.5cm,width=6.5cm,angle=0]{fig1.eps}\\
{\small{\bf Figure 1.}~
Frequency of occurence of different Quality values for an algorithm for a Travelling
Salesman Problem for number of nodes $N =$ 30, 50 and 70, over $10^6$ instances. The
distribution obeyed is eqn.(4), with $\alpha$, $\beta$ \& $\gamma$ values being
approximately given by: 72.57, 43.67, 255.45, respectively for $N = 30$;~~\& 242.72, 46.31, 425.31, 
respectively for $N = 50$;~~ and 317.69, 51.55, 617.56, respectively for $N = 70$}.\\

The frequency with which a certain quality $Q$ is achieved is shown in Figure 1 for a Travelling
Salesman tour with the number of nodes $N = 30, 50$, and $70$. The quality is found to obey
a distribution of the form
\begin{equation}
P(Q) = \alpha \exp(\beta Q - \gamma Q^3),
\end{equation}
$\alpha$, $\beta$ and $\gamma$ being $N$-dependent, positive parameters.\\
\noindent
The question thus arises -- how do we locate or predict a phase transition in a problem
such that it can be of practical import and used? To deal with this, we define a function $K$, which
we call the Cost Inflection function, the suddenly changing behaviour of which would indicate the
occurence of a phase transition. Since we want as good a solution as soon as possible, with minimal
time elapse, we could think of $K$ as the value of the solution corrected to
account for the time expended in arriving at it. We make the ansatz that the Cost Inflection function $K(Q_i,t_i)$
is related to the Value function $V(Q_i,t)$ and to solution quality through the relation:
\begin{equation}
K(Q_i,t) = V(Q_i,t) - t Q_i,
\end{equation}
where the value $V(Q_i,t)$ is defined through equation (3), $Q_i$ being the quality function of
the currently available solution.\\
\vspace*{0.65cm}\\
\noindent
\includegraphics[height=6.5cm,width=6.5cm,angle=0]{fig2.eps}
\includegraphics[height=6.5cm,width=10.5cm,angle=0]{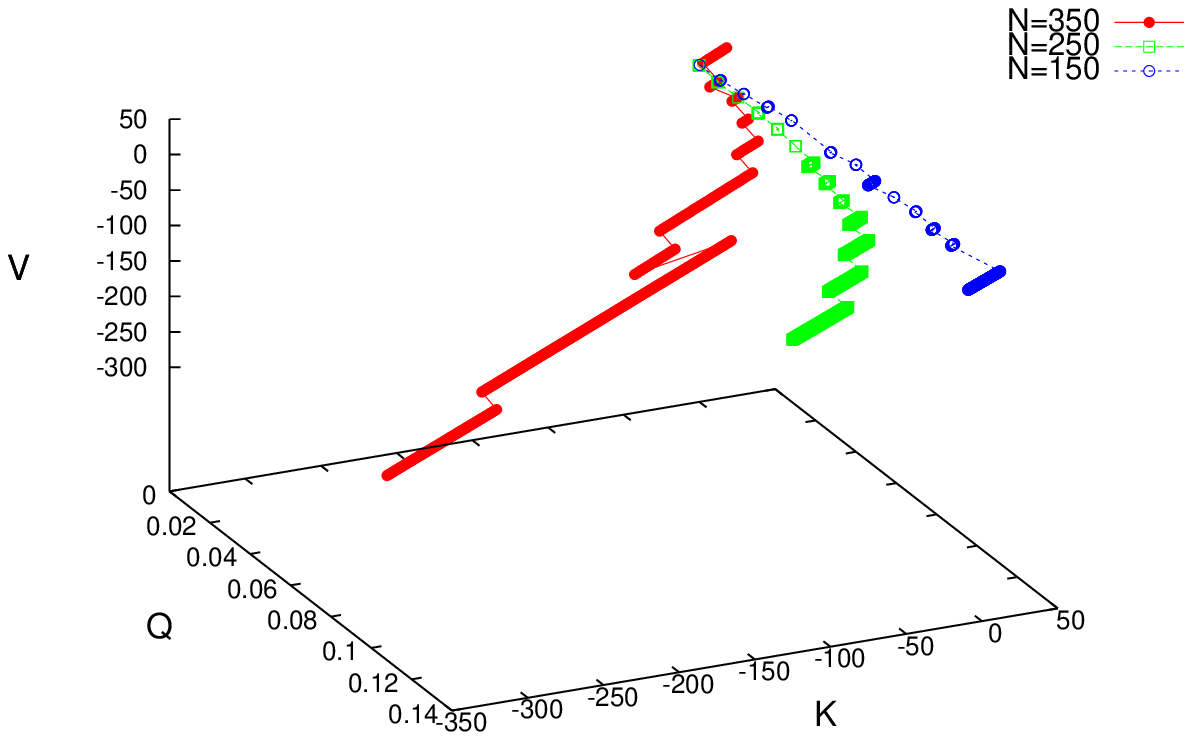}\\
{\small{\bf Figure 2.}~ Plot of the Value function $V$~~~~~~~{\bf Figure 3.}~Three dimensional plot of Cost-}\\ 
{\small as a function of time $t$, for different cases~~~~~~~Inflection, Quality and Value Functions,}\\
{\small of Travelling Salesman Problems with ~~~~~~~~~~~for $N =$ 150, 250 and 350.}\\
{\small number of nodes $N = 100, 250, 300$ and $350$.}\\

Figures 2 and 3 show the dependence between Cost Inflection, Value and Quality functions.
As can be clearly seen,
the Cost Inflection function shows a marked change in behaviour after a
particular point. We identify $K$ as an analogue to the Helmholtz free energy $A$.\\
\indent
Recall that the Value function is a weighted sum of the Utility function. The Value function's
temporal evolution (Figure 2) reminds us of the time evolution of the most probable value $X_m$ of the
number of particles (corresponding to the chemical composition) for a system in the stochastic
theory of adiabatic explosion~\cite{barasNicolisMalekmansourTurner}. In that thermodynamic system,
$X_m$ evolves with time till some critical time $t_c$ after which new solution-branches appear corresponding
to other probability states. In a system undergoing combustion, this is the temporary situation where the
molecules can be differentiated into a part for which combustion has not yet taken place and a part for
which combustion has ended. In our problem, the evolution of $V(f_i,t)$ is an evolution of the system to
higher quality solutions, differentiated by a point of inflection at a time $t = t_c$ where the
system slows down heralding the onset of critical slowing down.\\
The appearance of a symmetric spike in the ${\ddot V}$ profile plotted as a function of $t$
and $K$, and a sharp inflection of the curve into another plane, as shown in Figure 4, are
symptomatic of a major transition in quality, and is accompanied by a transition to a
different computational regime.\\
\vspace*{0.6cm}\\
\includegraphics[height=7.5cm,width=7.5cm,angle=0]{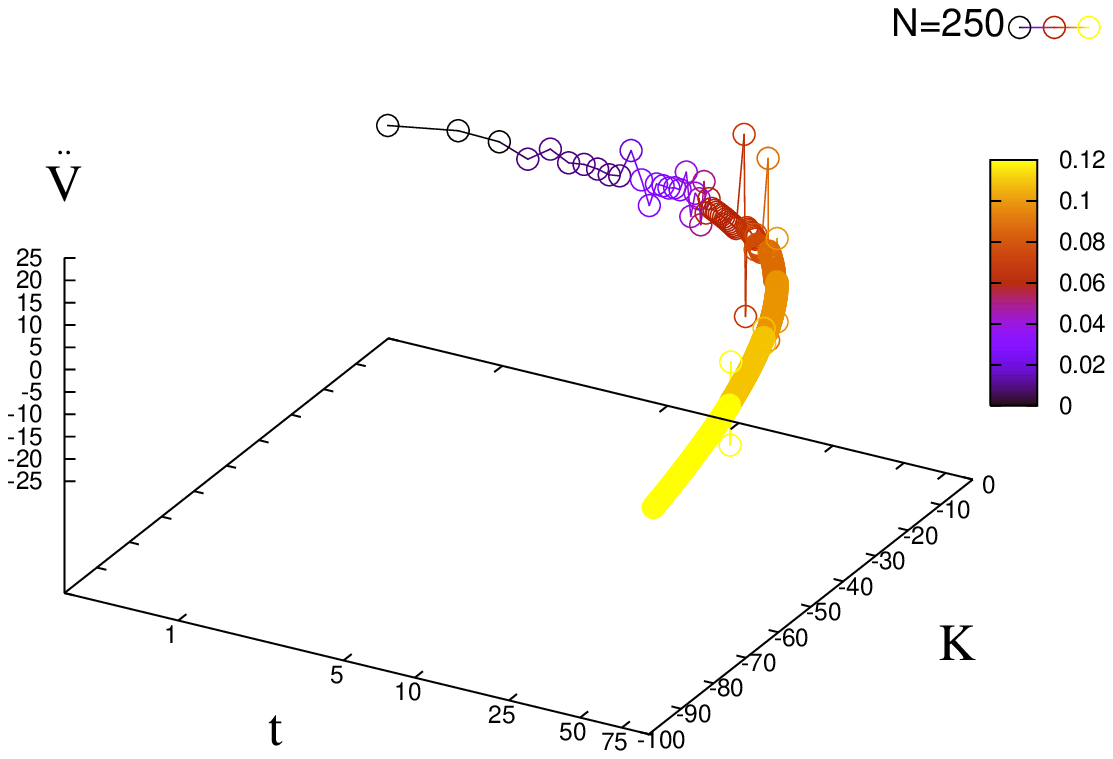}\hspace*{0.2cm}
\includegraphics[height=6.5cm,width=9cm,angle=0]{fig5.eps}\\
{\small {\bf Figure 4.}~The second derivative of the~~~~~~~~~~~{\bf Figure 5.} The Quality Function plotted as a}\\
{\small Value Function with respect to time, ${\ddot V}$~~~~~~~~~~~function of time for three values of $N$ ($N$ =}\\
{\small plotted as a function of $K$ and time $t$, the~~~~~~~~100, 200, 250), and an overlaying plot of an}\\
{\small Quality of the solution, $Q$, being depicted~~~~~~~~Arrhenius function, to show the broad}\\
{\small by the colour; number of nodes $N = 250$.~~~~~~~~~qualitative similarity.}\\
{\small Note the sharp inflection corresponding}\\
{\small to a phase transition.}\\

In Figure 5 we have plotted the improvement of solution quality with time for the TSP defined above,
for different values of $N$.
We find that the behaviour of $Q$ can be approximated by means of an Arrhenius equation
\begin{equation}
Q_{approx} = b \exp(-c/t),
\end{equation}
$b$ and $c$ being some real constants. The suffix {\it approx} has been added to $Q$ to stress that this is but
a tool in our gedankenexperiment to help understand a rather abstract system in terms more familiar to us.
This form of $Q_{approx}$ is almost exactly like the Arrhenius law for the rate constant $w(T)$ in a chemical
reaction in a combustive process $w(T) = A \exp (\frac{-E}{RT})$, $E$ being the activation energy,
$R$ the gas constant, $T$ the temperature, and $A$ the pre-exponential or frequency factor that can be
approximated to be constant.
Just as in the thermodynamic situation the rate constant increases rather steeply with temperature,
eventually plateauing out to a constant value, in our case the solution quality too increases quickly
with time before approaching the maximum achievable value for the problem.\\

\noindent
{\bf a}
\includegraphics[height=6cm,width=8cm,angle=0]{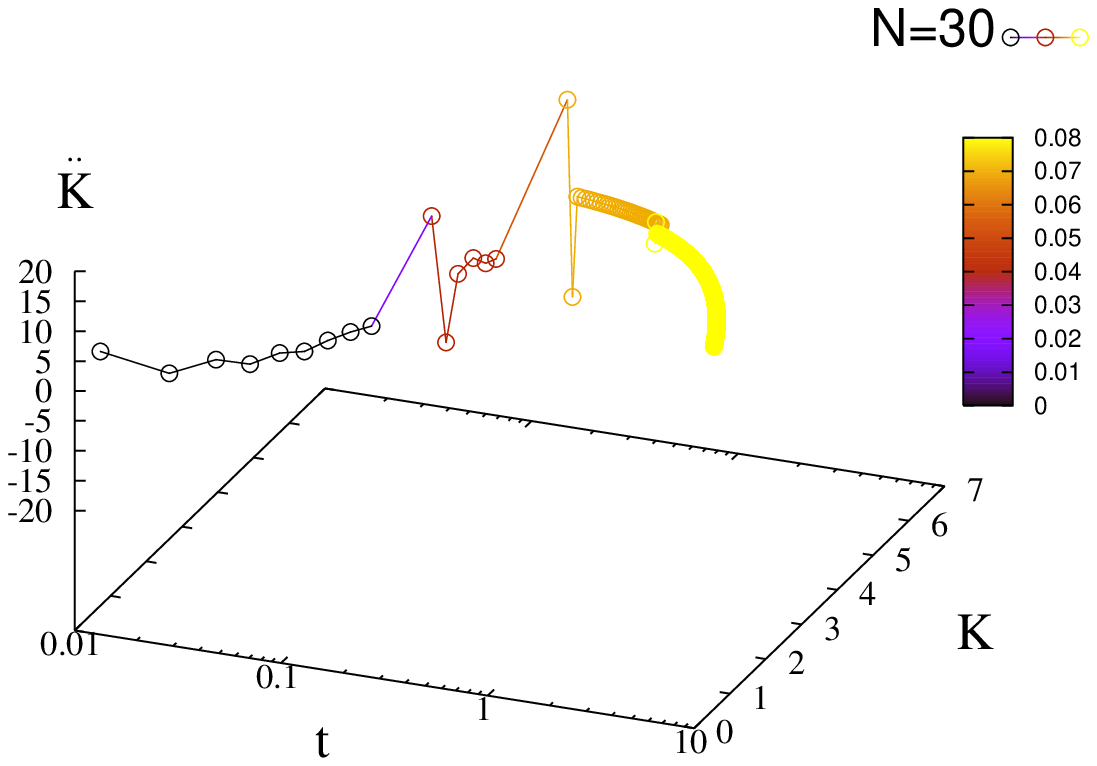}
\hspace*{-0.5cm}
{\bf b}
\includegraphics[height=6cm,width=8cm,angle=0]{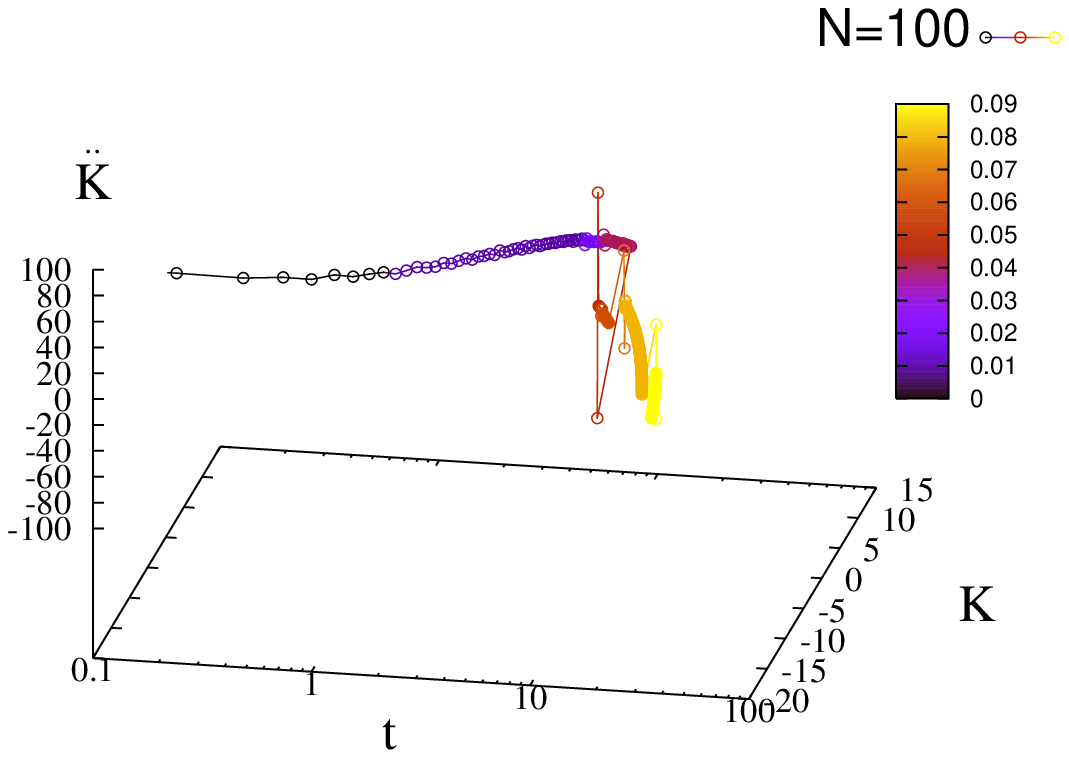}\\
\vspace*{1cm}\\
\noindent
{\bf c}
\includegraphics[height=6cm,width=8cm,angle=0]{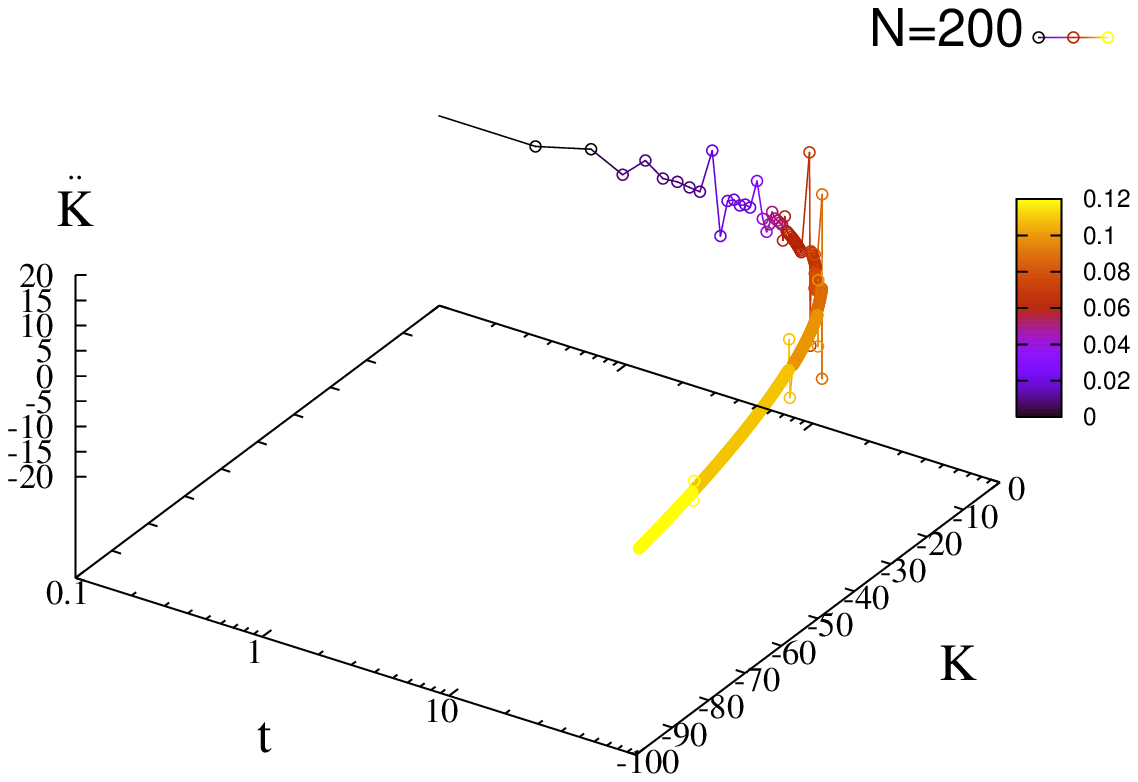}
\hspace*{-0.5cm}
{\bf d}
\includegraphics[height=6cm,width=8cm,angle=0]{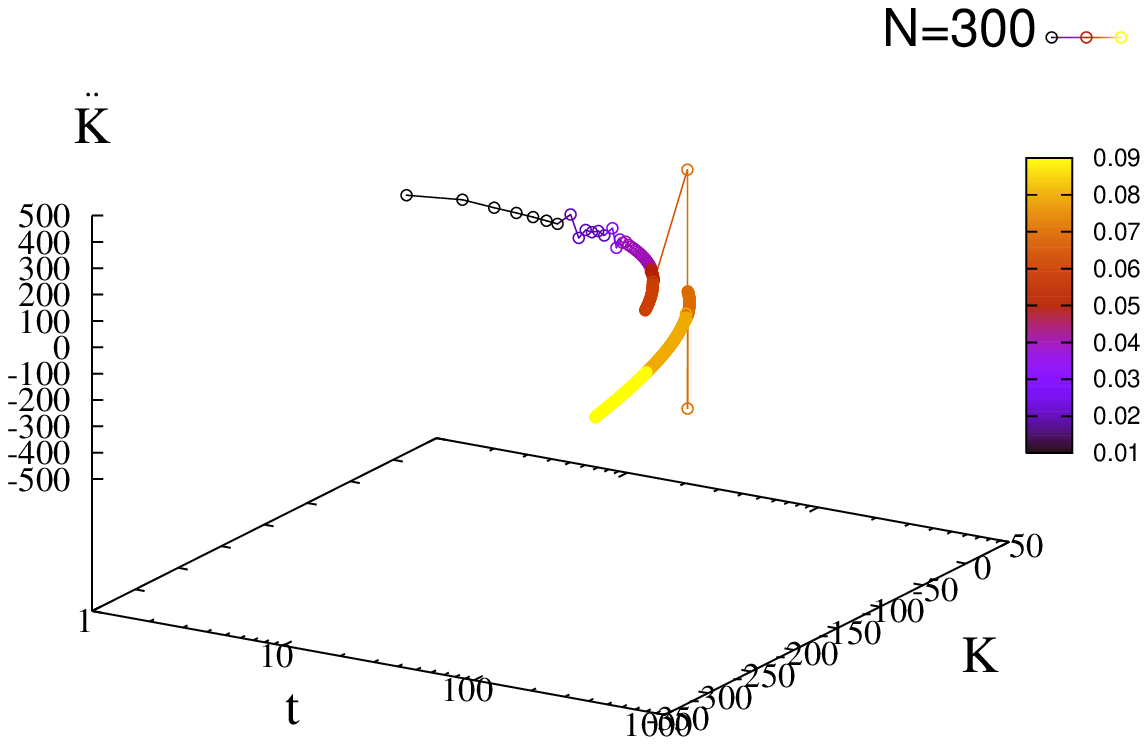}\\
{\small{\bf Figure 6.}~Representative $K-t-{\ddot K}-Q$ plots obtained for travelling salesman 
problems with $N=$ 30, $N=$ 100, $N=$ 200 \& $N=$ 300. $K$ is the Cost Inflection Function, 
$t$ the time, ${\ddot K}$ corresponds to the second derivative of $K$ with respect to time. 
$Q$, the Quality  function, is represented by the colour.}\\

As in thermodynamics, in our computational context too one can define appropriate response functions.
As time elapses, the system responds by a perceptible change in its quality. This is reflected as a
decrease in the cost inflection function $K$ after a critical value (see Figure 6). This motivates us to
define a quantity ${\mathcal C}_N$, for which we coin the term efficacy of the algorithm:
\begin{equation}
{\mathcal C}_N = |t {\ddot K}|,
\end{equation}
where derivatives are with respect to time. The cost inflection function $K$ is calculated from
equation (5).  A plot of the efficacy function, ${\mathcal C}_N$, as a function of time $t$ for different
$N$ values (Figure 7 a-b) shows an interesting observation -- ${\mathcal C}_N$ tends to show
almost random behaviour upto a point where a transition to a stable regime takes place;
thereafter, the efficacy function shows a strict $t$ dependence
\begin{equation}
{{\mathcal C_N}_{stable}} \propto t.
\end{equation}
\noindent
{\bf a}\\
\begin{center}
\includegraphics[height=8.5cm,width=12.75cm,angle=0]{fig7a.eps}\\
\end{center}
\noindent
{\bf b}\\
\begin{center}
\includegraphics[height=5cm,width=7.2cm,angle=0]{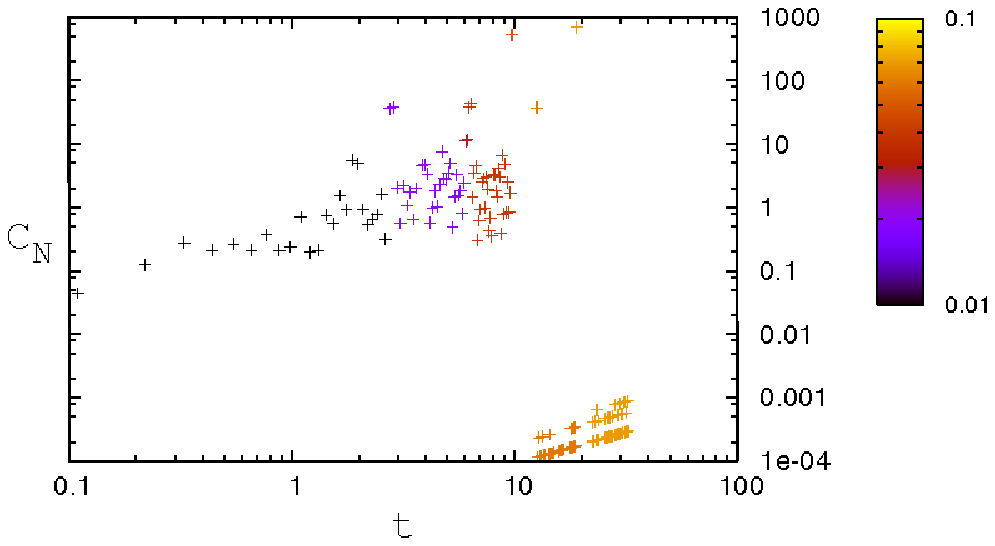}
\includegraphics[height=5cm,width=7.2cm,angle=0]{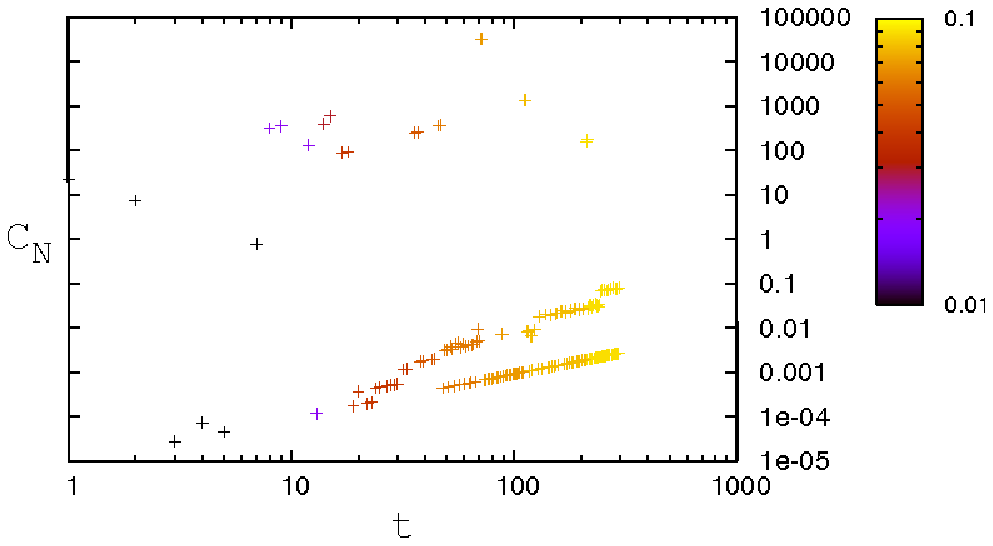}
\end{center}

{\bf Figure 7.}~Efficacy ${\cal C}_N$ plotted as a function
of time $t$ (a) for different number of nodes $N$, all showing a transition to a stable regime where
${{\cal C}_N}_{stable} \propto t^2$; (b) for $N = 175$ at left, and $N=300$ at right: representative plots showing the regime of
${\cal C}_N = {{\cal C}_N}_{stable}$ to correspond to the high quality metastable \& stable states.\\

This stable area, as can be seen in Figure 7b, corresponds to regimes where the quality
$Q$ has reached its maximal value, or is reaching it, and encompasses, therefore, both
stable and metastable states. The efficacy as defined above therefore makes a
very good response function in identifying a transition to a near-optimal set of regimes.\\
In the thermodynamic context, the specific heat $C_v$ for a fluid is, as we know, defined by
\begin{equation}
C_v = - T \left(\frac{\partial^2 A}{\partial T^2}\right)_v,
\end{equation}
where $A$ denotes the Helmholtz free energy. The specific heat behaviour as a function of
temperature is distinct, depending upon whether the temperature is below or above a critical
temperature $T_c$ for the thermodynamic system. For example, for the well-known superfluid
or $\lambda$ transition in $^4$He the specific heat dependence can be approximated by
$C \approx A~\log(T - T_\lambda) + B$, for temperatures above the critical $\lambda$-transition
temperature $T > T_\lambda$, and $C \approx A'~\log(T_\lambda - T) + B'$ for temperatures below the
critical temperature $T < T_\lambda$, with the prefactors $A$, $B$, $A'$ and $B'$ being
temperature - independent constants~\cite{goldenfeld}.\\
In our case, the response function ${\mathcal C}_N$ is such that its behaviour changes abruptly
at the point corresponding to a critical value of elapsed time for the system.
In other words, looking at ${\cal C}_N$, we should clearly be able to distinguish between separate
phases or regimes.

Approach to a phase transition or bifurcation point in physical systems is typically
characterized by a critical slowing down. We therefore expect that our system should
likewise show similar behaviour. In the Travelling Salesman Problem under consideration,
critical slowing down would correspond to a regime wherein solution quality remains
unchanged over a comparatively long period of time, while other measures of algorithm
efficiency, like the cost inflection function $K$, or the value function $V$, tend to
show a change in behaviour.
An investigation of our data indicates that the onset of critical slowing down in the
algorithm can be quantified and detected. Locating $t_c$, the time of onset of this critical 
slowing down, is quite straightforward.\\
\vspace*{0.6cm}\\
\includegraphics[height=6.5cm,width=6.5cm,angle=0]{fig8.eps}\\
{\small {\bf Figure 8.}~Plot indicating dependence of $t_c$, time of onset of critical 
slowing down, on number of nodes $N$. The best fit for the log-log plot is the equation
$t_c = a_0 N^z$, where $a0 = 0.00047 \pm 0.00027$, $z = 2.073 \pm 0.099$.}\\

\indent
It should be noted that there may be more than one regime where the solution quality
of the algorithm lingers at one value for an extended period of time.
Such states may essentially be thought of as metastable states for the system, as has
been mentioned in our discussion of the efficacy function above.
$t_c$, however, is distinguished from such metastable states by the clearly
discernable change in behaviour of functions like $K$, $V$ or ${\ddot V}$.
We find that a
plot of $t_c$, the time corresponding to the onset of critical slowing down,
versus $N$, the number of cities or nodes
in a 2-D Travelling Salesman Problem, (Figure 8), scales like:
\begin{equation}
t_c \sim N^{z}, z \approx 2.07.
\end{equation}
This critical exponent is reminiscent of the dynamical scaling exponent $z$,
in physical systems~\cite{stanley,goldenfeld,chaikin}.
We note that our system is intrinsically non-autonomous.
Figure 6 shows representative plots of $K$ plotted against ${\ddot K}$ and time, with the
Quality of the solution being represented by the colour. This is yet another way of looking
at how there is a change in the evolution of the Cost Inflection function with time.\\
The presence of large fluctuations can be noted in the plots shown in Figure 6 and, for example,
in Figure 4. Apart from the fact that such fluctuations would be expected as one approaches
a point of bifurcation, Figure 7a suggests too that for a given value of $N$, the system can
have more than one, close ``energy state'', so that the system tends to fall into either one
or more of these metastable and stable states during the course of its evolution.
Generic theoretical treatments of fluctuations near bifurcations that correspond to the
appearance of new stable states have appeared in the literature (for example,
Reference~\cite{dykman}).
However, an analysis of the fluctuations observed in our system and the apparent absence of
correlation of fluctuation size to system size, is outside the scope of this present work,
and will not be dealt with here.\\

\noindent
Differentiating eqn.(5) twice with respect to time,
\begin{equation}
t{\ddot Q} + 2 {\dot Q} = {\ddot V} - {\ddot K} .
\end{equation}
Since ${\ddot K} - {\ddot V}$ is very small and near zero for most times after the critical point
corresponding to the time $t_c$, beyond which we can justifiably expect that no further temporal
evolution of the rate at which the Value and Cost Inflection functions change with time takes place, we can write
\begin{equation}
t{\ddot Q} + 2 {\dot Q} \approx 0.
\end{equation}
Integrating this between the limits $t = t_a$ and $t_b$, $t_b > t_a$, we get
\begin{equation}
Q_b - Q_a = k (\frac{1}{t_a} - \frac{1}{t_b}),
\end{equation}
$k$ being some constant of integration.
If we now identify $t_a$ as the time corresponding to critical slowing down, $t_c$,
and take $t_b$ to be much larger, so that $t_b \rightarrow \infty$, we are left with
\begin{equation}
Q_{max} - Q_c \approx k/t_c,
\end{equation}
where $Q_c$ is the value of the quality function at the onset of critical slowing down
and $Q_{max}$ is the value when the algorithm is run for a much longer time. It will be appreciated
that equations (13) and (14) are valid only for time $t \geq t_c$. \\
As we have seen (Figure 5, equation (6)), the variation of $Q$ with time $t$ can be very broadly
approximated by an Arrhenius-like functional dependence on time $Q = b\exp(-c/t)$. At large times,
this can be approximated to $Q \approx b (1 - c/t)$, so that
\begin{equation}
Q_{max} - Q_c \approx ~b ~c /t_c.
\end{equation}
We are therefore able to give a value to the constant $k$ of equation (14),
\begin{equation}
k \approx b c,
\end{equation}
so that the solution quality at $t_c$ is readily predictable:
\begin{eqnarray}
Q_c &=& Q_{max} - b~c~/ t_c \nonumber \\
&\approx& b(1 - c/t_c),
\end{eqnarray}
since $Q_{max} \approx b$.\\

\noindent
An order parameter $m_q$ for this time-evolving system could be a measure of quality defined by
\begin{equation}
m_q = Q_c - k(1/t - 1/t_c).
\end{equation}
For $t \geq t_c$, $m_q$ would necessarily be greater than equal to $Q_c$, i.e., $m_q - Q_c \geq 0$.
For any time $t < t_c$, $m_q$ would be less than $Q_c$, so that $m_q - Q_c < 0$.\\

Another issue of interest in travelling salesman problems is obtaining an estimate of the
optimal tour length. The optimal tour distance $L_o$ for a $d$-dimensional TSP is a function of $d$
and the number of cities $N$. In the large $N$ limit, this optimal tour distance is given
by~\cite{beardwoodhaltonhammersley}
\begin{equation}
\lim_{N \to \infty}\frac{L_o}{N^{1-1/d}} = \alpha_o,
\end{equation}
$\alpha_o$ being a constant.
Percus and Martin have investigated the finite size scaling of the optimal tour length of this
Euclidean TSP with randomly distributed cities~\cite{percusmartin}. For the two-dimensional case,
they found that $L_o$ can be written in the form
\begin{equation}
\frac{L_o}{N^{1/2} (1 + 1/(8N) + ...)} = \alpha_o(1 - 0.0171/N + ...),
\end{equation}
with $\alpha_o = 0.7120$~\cite{percusmartin}. In an earlier paper reporting the simulated annealing
simulations of a travelling salesman problem~\cite{leechoi}, Lee and Choi obtained the optimal tour
length to be given by
\begin{equation}
L_o(N) = \alpha \sqrt{N} +\beta,
\end{equation}
$\alpha \equiv \alpha(N \rightarrow \infty)$, $\alpha(N) \equiv L_o/\sqrt{N}$, getting an upper bound
value of $\alpha = 0.7211$, $\beta = 0.604$.\\
It would also be interesting to know how close to optimality are the solutions that we have
obtained from our algorithm. Keeping in mind that what we seek to do in this paper is to describe
a technique for deciding when to halt a computation while approaching a solution that is approaching
optimality, rather than obtaining optimum solutions, we find from our data that for our best solutions,
\begin{equation}
L_o (N) \approx 0.7408 \sqrt{N} + 1.1732
\end{equation}
holds, i.e., $\alpha = 0.7408$, and $\beta = 1.1732$.
Figure 9 shows this in a plot of the length of the optimal solution $L_o$ versus $1/\sqrt{N}$.\\
\vspace*{0.6cm}\\
\includegraphics[height=6.5cm,width=6.5cm,angle=0]{fig9.eps}\\
{\small {\bf Figure 9.} Plot of optimal length $L_o$ vs. $1/\sqrt{N}$. The plot obeys equation (22), 
~$L_o(N) = 0.7408\sqrt{N} + 1.1732$.}\\

The general trend of our solutions, therefore, are consistent with that in the literature. Improved
values for $\alpha$ can more easily be obtained when the system size is much larger, and when computations
are run longer.

\section*{\large{\bf 5. Discussion and Conclusions}}
From the above treatment, one can see that it would be possible to look out
for drastic transitions in search costs and solution quality while monitoring
algorithms. In the case of Anytime Algorithms, for example, we could now effectively
identify transitions from
contract to interruptible algorithms. Identifying the transition and the associated behaviour,
required us of course, to run our algorithm for a long period of time, irrespective of
solution quality.
It should be noted that the quality measure could be defined in several ways, and we have
chosen but one possible definition. Our choice of obtaining an initial configuration
using a greedy assignment of nearest neighbour jumps was made because it works well for a
random 2D Travelling Salesman Problem, which was the toy problem we chose for testing
our ideas. This need not, of course, be used for other problems.\\
\indent
Our method does not set out to explicitly find a global minimum corresponding to some
particular stable state, but rather, to locate the existence of both stable and
metastable states. The intention is to be able to find and predict regimes where
computational time spent is not commensurate with improvement in solution quality achieved.

\indent
We have defined a response function, somewhat akin to a specific heat of a fluid,
whose behaviour handily identifies metastable and stable states corresponding to high
solution qualities. This efficacy ${\cal C}_N$ is a clear indicator of a
transition from lesser quality unstable solutions to the higher quality stable solutions,
showing a $t$-dependence for the latter, for the TSP problem we have considered.

We have identified a dynamical critical exponent $z \approx 2.07$ relating the onset
of critical slowing down place to the number of nodes in a travelling salesman problem.
This is a very invaluable result: we can now know when a transition in quality will occur and
have a handle on how long an algorithm need be run. It will be recalled that all possible
rearrangements in a 2-opt algorithm are of order $N^2$. The coincidence of this with
our critical exponent of 2.07 underlines the fact that our approach is robust and
is consistent with what would be expected from a more traditional approach.
It is crucial, however, to note that it is not the enumeration of $N^2$ 2-opt states
that we have called a dynamical critical phenomenon, but rather the onset of critical
slowing down, since time is indeed a control parameter with all our functions (quality,
cost, utility, response function, etc.) varying dynamically in time.
We expect that similar universal exponents would be valid for other examples
in this class of problems.\\
\indent
 It becomes meaningful therefore to conduct similar investigations on
other problems and look for similar critical exponents. Further work on this is
underway and will be reported elsewhere.\\
While we have implemented our formalism and demonstrated its working using a 2-opt algorithm for
a travelling salesman problem, this generic form may be suitably modified for use on any other
problem using any other algorithm. This is the real advantage of our work -- it enables one to
set the time parameters under which an anytime algorithm may be used as a contract anytime algorithm,
i.e., with the time of its running being fixed in advance; we are also able to quantify
an abstract quantity like solution quality in a generic fashion by means of a response function which
clearly indicates the onset of a phase transition in solution quality and hardness.

\section*{\large{\bf 6. Acknowledgements}}
We would like to thank Profs. Subhash Chaturvedi and J. Balakrishnan for useful discussions
and suggestions, and the anonymous referee for useful suggestions on improving the manuscript.

\end{document}